\documentclass[useAMS,usenatbib]{mn2e}
\usepackage{times}
\usepackage{aas_macros}
\usepackage{amssymb}
\usepackage{graphicx}
\title[The accretion rate and minimum spin period of accreting pulsars]
      {The accretion rate and minimum spin period of accreting pulsars} 
\author[A. Bonanno and V. Urpin]
  {A.~Bonanno$^{1,2}$ and V.~Urpin$^{1,3,4}$ \\
      $^{1}$ INAF, Osservatorio Astrofisico di Catania,
      Via S.Sofia 78, 95123 Catania, Italy \\
      $^{2}$ INFN, Sezione di Catania, Via S.Sofia 72,
      95123 Catania, Italy \\
      $^{3}$ A.F.Ioffe Institute of Physics and Technology,
      194021 St. Petersburg, Russia \\
      $^{4)}$ Isaac Newton Institute of Chile, Branch in St. Petersburg,
           194021 St. Petersburg, Russia}
\date{today}

\pagerange{\pageref{firstpage}--\pageref{lastpage}} \pubyear{2002}

\def\LaTeX{L\kern-.36em\raise.3ex\hbox{a}\kern-.15em
    T\kern-.1667em\lower.7ex\hbox{E}\kern-.125emX}

\begin{document}

\label{firstpage}

\maketitle

\begin{abstract}
We consider combined rotational, magnetic, and thermal evolution
of the neutron star during the accretion phase in a binary system.
A rapid accretion-driven decay of the magnetic field decreases
substantially the efficiency of angular momentum transfer.
As a result, the neutron star cannot spin up to periods shorter
than some limiting value even if accretion is very long and 
accretion rate is high. The proposed mechanism can explain
a discrepancy between the shortest detected period and 
minimal possible spin period of neutron stars.  

\end{abstract}

\begin{keywords}
magnetic field  - stars: neutron - pulsars: general - stars: accretion -
stars: rotation

\end{keywords}

\section{Introduction}  

According to the generally accepted point of view \citep{bhatta91}
the population of radio millisecond pulsars 
originates from low mass X-ray binary systems (LMXBs). The recycling 
scenario suggests that a neutron star born in a binary system with more 
or less typical parameters can accrete the gas stripped by the companion 
until its magnetic field and spin period will fit the so-called 
"millisecond pulsar box" in the $B$-$P$ (magnetic field - period) plane. 
The characteristic values of the magnetic field and period for 
millisecond pulsars are $B \sim 10^8 - 10^9$ G and $P \sim 1-10$ ms,
respectively, and these values can be reached as a result of evolution
in a binary system. The evolution of a neutron star in a low-mass binary 
is extremely long and complex, because even the main-sequence lifetime 
of a low-mass companion exceeds $10^9$ yrs. In accordance with the 
standard scenario \citep{pringle72,Illarionov75}
the neutron star in a close binary passes throughout several 
evolutionary phases: \\
I) The initial phase, in which the pressure of the pulsar radiation
keeps the wind plasma of a companion away from the neutron star
magnetosphere. The magnetic, thermal, and rotational evolution do
not differ from those of an isolated star. \\
II) The propeller phase, in which the radiation pressure reduced by
the spin-down and field decay cannot prevent the wind plasma from 
interaction with the magnetosphere. However, rotation is still 
sufficiently fast to eject the wind plasma by a propeller effect. \\
III) The wind accretion phase, in which the wind plasma falls down
on to the surface of a neutron star and nuclear burning of the accreted 
material heats the star and, due to this, accelerates the field 
decay. \\
IV) The enhanced accretion phase, which starts when the companion
leaves the main-sequence and fills its Roche lobe. The Roche-lobe
overflow drastically increases accretion on to the neutron star.
During this phase, a vigorous mass transfer heats the neutron star 
interiors to a very high temperature, $(1-3) \times 10^8$ K and 
accelerates essentially the field decay. The accretion torque spins 
up the neutron star to a short period.

The pulsars processed in the above transformations can have 
finally the parameters close to those of millisecond pulsars 
\citep{urpin98aa,urpin98mn}
even if they had initially the typical 
for pulsars magnetic fields and periods. Certainly, the neutron star 
experiences  the most dramatic changes in the course of the phase 
IV, and this phase determine the final state of the pulsar. 
During the enhanced accretion phase, the neutron star initially 
spins up very rapidly until it approaches the so called spin-up 
line corresponding to the accretion rate. 
The spin-up line in the 
B-P plane is determined by corotation at the Alfv\'en radius. It 
was suggested by \cite{ba91} that, during 
the further evolution, a balance between spin-up and the rate of 
field decay is reached, so that the neutron star slides down the 
corresponding spin-up line with a rate that is determined by the 
field decay. 

However, the behavior of a neutron star during the enhanced 
accretion phase can be much more complicated if one takes into 
account  the combined magnetic, thermal, and rotational evolution.
Because of a high internal temperature, 
the magnetic field decay during the accretion phase can be very 
fast such as mass transfer is not able to provide a sufficient 
amount of the angular momentum to maintain a balance at the spin-up
line. Therefore, a significant departure from the spin-up line 
can be expected during the accretion phase. In this paper, we will
show that these departures occur at both low and high accretion 
rates and, owing to them, the minimum period exists that can limit 
spin-up of the accreting neutron star. The present study is 
addressed mainly the neutron stars in low-mass X-ray binaries 
(LMXBs). However, the same qualitative behavior is typical for 
spin-up of accreting neutron stars in other types of binary 
systems as well. This concerns particularly high-mass X-ray 
binaries where the accretion rate is very high and field decay
is fast \citep{urpin98hm}. In such systems, 
the angular momentum transfer can also be substantially influenced
by the magnetic field decay and thermal evolution. The considered 
mechanism of spin-up can be the key issue in understanding the 
origin of various classes of accreting pulsars.

\section{Basic equations}

Consider the evolution of a neutron star during the enhanced 
accretion phase that begins at the end of the main-sequence life 
of a companion. The magnetic, thermal. and spin evolution of the 
neutron star may be essentially affected by such accretion. We 
assume that the magnetic field is maintained by electric currents 
in the crust, Such magnetic configuration can be generated, for 
example, by turbulent dynamo during first minutes of the neutron 
star life \citep{bonanno05,bonanno06}. The evolution of the crustal 
field is determined by the conductive properties of the crust and 
material motion throughout it. In a very strong magnetic field, 
the ohmic dissipation can be accompanied also by non-dissipative 
Hall currents. These currents affect the decay of the magnetic 
field indirectly, coupling different modes and generating 
magnetic features with a smaller length-scale than the background 
magnetic field \citep{naito94}. Numerical simulations 
\citep{hr02,hr04} indicate that some 
acceleration of the decay of a large scale field (for instance, 
dipole) can occur if the Hall parameter is very large but the 
effect is not significant for typical pulsar fields. Besides,
this effect turns to be sensitive to the initial magnetic geometry. 
For example, the presence of both the toroidal and dipole field 
components decreases the rate of dissipation caused by the Hall
effect, and dissipation proceeds approximately on the ohmic 
time scale \citep{hr02,hr04}. Note also
that the Hall currents are important only for a very strong 
magnetic fields but, at the beginning of enhanced accretion, 
the field of a neutron star is usually weaker than the "standard" 
pulsar field \citep{urpin98aa,urpin98mn}.
Therefore, we 
will neglect the Hall currents. Then, the induction equation in 
the crust reads 
\begin{equation}
\frac{\partial \vec{B}}{\partial t} =  - \frac{c^2}{4 \pi} \nabla
\times \left( \frac{1}{\sigma} \nabla \times \vec{B} \right) + 
\nabla \times ( \vec{v} \times \vec{B}),
\end{equation}
where $\sigma$ is the conductivity and $\vec{v}$ is the velocity 
of crustal matter. The velocity $\vec{v}$ is caused by the flux 
of the accreted matter and is non-zero only during the periods 
when the neutron star undergoes accretion. This flux carries 
out the magnetic field into the deep crustal layers. Assuming 
spherical symmetry of the material flow, the velocity in the 
negative radial direction can be written as
\begin{equation}
v = \frac{\dot{M}}{4 \pi r^2 \rho},
\end{equation}
where $\dot{M}$ is the accretion rate and $\rho$ is the density.
We consider the evolution of an axisymmetric magnetic field
following \cite{wendell87}. Introducing the vector potential 
$\vec{A} = (0, 0, A_{\varphi})$ where $A_{\varphi} = S(r, \theta, 
t)/r$ and $(r, \theta, \varphi)$ are the spherical coordinates, 
we obtain the equation for $S$ from Eq.(1).
The function $S$ can be separated in $r$ and $\theta$ in the form
\begin{equation}
S = \sum_{l=1}^{\infty} s_l (r, t) P_l^1 (\cos \theta),
\end{equation}
where $P_l^1(\cos \theta)$ is the associated Legendre polynomial 
with the index 1.  Using the properties of Legendre polynomials, 
we have
\begin{equation}
\frac{\partial s_l}{\partial t} = v \frac{\partial s_l}{\partial r} +
\frac{c^2}{4 \pi \sigma} \left[ \frac{\partial^2 s_l}{\partial r^2} -
\frac{l(l+1)}{r^2} s_l \right].
\end{equation}
In the case of a dipole field ($l=1$), the function $s_1(r, t)$ can
be related to the surface magnetic field at the pole, $B_p(t)$, by
$B_p(t) = 2 s_1 (R, t)/ R^2$ where $R$ is the stellar radius 
\citep{urpin94}.
Since Eq.~(5) is linear, we can 
normalize $s_1$ in such a way that $s_1(R, 0)=1$. Then, the ratio 
$B_p(t)/ B_p(0)$ is given by $s_1(R, t)$.  

Continuity of the magnetic field at the stellar surface $r=R$ yields 
the boundary condition for Eq.(5)
\begin{equation}
\frac{\partial s_l}{\partial r} + \frac{l}{R} s_l = 0.
\end{equation}
In this paper, we consider the case of a dipolar field with $l=1$.

Since the crustal conductivity depends on the temperature $T$, 
accretion must influence the field decay by changing both $T$ 
and inward directed flux of matter. Therefore, the magnetic 
evolution of a neutron star in a binary turns out to be strongly 
coupled to its thermal evolution.
The thermal evolution is determined mainly by pycnonuclear
reactions and, due to them, the crust is heated by nuclear 
burning of the accreted material. The enhanced accretion heats 
the neutron star to a high temperature  $\sim (1-3) \times 10^8$ K 
that leads to a rapid field decay. The thermal structure of the 
star with pycnonuclear reactions has  been studied by a number 
of authors \citep[and references therein]{brown98,zdu08}. 
In LMXBs, accretion due to Roche-lobe 
overflow can last as long as $10^6 - 10^8$ yrs and the magnetic 
field can be drastically reduced in the course of this phase. 

{Our model is in contrast to the widely accepted assumption
of a proportionality between  to the amount of accreted material 
and the field decay rate at any time along the spinup line \citep{shiba}. 
This simple model 
is obviously inconsistent from a theoretical point of view because 
the decay is determined not only by the duration of accretion (or, 
equivalently, by the total amount of accreted mass) but also by the 
conductivity of the crust. The latter is determined by the temperature 
which is dependent on the accretion rate. As a result, a consistent 
description of the field decay must take into account both the 
duration of accretion (or total amount of accreted mass) and accretion 
rate. A detailed study of the accretion-driven field decay \citep{ug}
reveals that, generally, the decay is not proportional to the amount 
of accreted mass at any accretion rate. Departures from this simple 
relation are essential even at a low accretion rate but the dependence 
becomes particularly complicated if the accretion rate is high, 
$\geq 10^{-10}$ $M_{\odot}$/yr (see, e.g., Fig.~5 of the paper by 
\cite{ug}). It is worth noting that a simple model with a field decay 
proportional to the amount of accreted material is not in agreement 
with observational data as well. This point is discussed in detail 
by \citep{wije}. In the model developed in the present study, the 
magnetic field decay is determined by both the accretion rate and 
duration of accretion.         
}

The rotational evolution is entirely determined by accretion as well
because the mass flow carries a large amount of the angular momentum 
to the neutron star. It is widely believed that the mass flow 
interacts with the magnetosphere at the so-called Alfven radius,
$R_A$, which is determined by a balance of the magnetic pressure 
and the dynamical pressure of plasma,
\begin{equation}
R_A = \left( \frac{2 R^6 B_p^2}{4 \dot{M} \sqrt{GM}} \right)^{2/7};
\end{equation}  
where $M$ is the neutron star mass and $G$ is the gravitational 
constant. If the neutron star rotates rapidly and its angular 
velocity is larger than the Keplerian angular velocity at the 
Alfven radius, $\Omega > \Omega_K(R_A) = (G M / R_A^3)^{1/2}$, 
plasma penetrating to the Alfven radius should be provided 
some portion of the angular momentum from the rapidly rotating 
magnetosphere and, as a result, this plasma must be expelled. 
On the contrary, matter carrying the angular momentum can falls 
down on to the neutron star if its rotation is relatively slow 
and $\Omega < \Omega_K(R_A)$. Due to this angular momentum, the 
neutron star spins up to a shorter period.  However, the star 
cannot spin up to a period shorter than the Keplerian period at 
the Alfven radius because otherwise it will work as a propeller 
and expels plasma instead of accreting it. It has been argued 
by \cite{ba91} that a balance should be 
reached in spin up and the rate of the field decay, thus the 
neutron star slides down the spin up line with the corresponding 
accretion rate. 

Since the interaction with the magnetosphere occurs at the Alfv\'en 
radius, the angular momentum carried by the accreted plasma can be
characterized by its Keplerian value at $R_A$, $\Omega_K(R_A) R_A^2$, 
multiplied by some "efficiency" factor $\xi$, $\xi < 1$. This factor 
depends on the geometry of accretion flow. If the accreted matter 
forms an accretion disk around the neutron star then $\xi \sim 1$. 
If accretion is close to spherical one, then $\xi$ is likely much
smaller. The rate of the angular momentum transfer to the neutron 
star, $\dot{J}$, can be estimated as $\sim \xi \dot{M}(\Omega_K(R_A) 
R_A^2) $, The corresponding spin-up rate is
\begin{equation}
\dot{P} \sim - \frac{P^2 \dot{J}}{2 \pi I} \approx - \xi \beta P^2 
B_p^{2/7} \dot{M}_G  ^{6/7},
\end{equation}  
where $\beta = (G M R^2/ 4)^{3/7}/ \pi I$, $I$ is the moment of 
inertia of the neutron star. The star spins up until the angular
velocity of the neutron star $\Omega$ becomes comparable to 
$\Omega_K(R_A)$. This condition determines the critical period,
$P_{eq}$, which distinguishes the accretion and propeller phases. 
We have for $P_{eq}$
\begin{equation}
P_{eq} \approx \frac{12.4 B_{12}^{6/7} R_{6}^{18/7}}{\dot{M}_{-10}^{3/7}
M^{5/7}_{1.4}} \; {\rm s},
\end{equation}
where $B_{12}= B_p/10^{12}$ G, $\dot{M}_{-10} = \dot{M} /10^{-10}
M_{\odot}$yr$^{-1}$, $M_{1.4}= M/ 1.4M_{\odot}$ and $R_{6}= R/10$km. 
Eq.(8) determines the
so-called spin-up line in the $B-P$ plane. Likely, the accreted 
matter forms the Keplerian disk during this phase. Therefore, spin 
up of the neutron star is driven by Eq.(7) with $\xi \approx 1$. 
At the beginning of the accretion phase, the neutron star approaches 
the spin-up line corresponding to an enhanced accretion rate and,
then, it slides down this line. Evolution on the spin-up line can 
last until either the accretion regime is changed or the magnetic 
field becomes too weak to maintain a balance in spin up and field 
decay. In the latter case, the neutron star must leave the spin 
up line and evolve in a very particular way. We assume that the 
spin period of a neutron star follows Eq.~(7) if $P \geq P_{eq}$. 
However, as it was discussed above, we have to suppose $P = 
P_{eq}$ if Eq.~(7) leads to $P < P_{eq}$ because accretion must
stop at such rotation and the neutron star begins to work as a 
propeller if $P < P_{eq}$, 
As a  result of accretion, we obtain a rapidly rotating neutron 
star with a low magnetic field that can manifest itself as a 
radiopulsar after accretion is exhausted.

\section{Numerical results}

The thermal evolution of a neutron star is considered by
making use of the public code developed by Page, {\it NSCool} 
\footnote{http://www.astroscu.unam.mx/neutrones/NSCool/}
in which  the magnetothermal evolution has been taken into 
account by coupling the induction equation with the thermal 
equation via the Joule heating as discussed by 
\cite{boba}.
In particular the induction equation has been solved 
via an implicit scheme and the electron conductivity has been 
calculated with the approach described \cite{pote99}
\footnote{http://ioffe.ru/astro/conduct}. 
Actual calculations have been performed for a neutron star 
of $M = 1.4 M_{\odot}$ based on the APR equation of state (EOS)
\citep{apr}. The radius and the thickness of the crust 
for this star are 11.5 km and $\approx 1.0$ km, respectively.
For the chemical composition of the crust, we use the same 
so-called ``accreted matter'' model by \cite{zdu08}. 
{In addition we have also considered models of $M = 1.4 M_{\odot}$
and $M = 1.6 M_{\odot}$
obtained with a stiffer  EOS where the barionic
matter is calculated using a field-theoretical models at the mean
field level \citep{gl}.
In particular the extension of the crust for $M=1.4 M_{\odot} $ for this EOS is about 2.1 km with  a radius of 13.8 km, 
at variance with the APR EOS case.}

The impurity parameter, $Q$, is assumed to be constant 
throughout the crust and equal to 0.001. The spin period 
at the beginning of the enhanced accretion phase is assumed 
to be $P_0 = 100$ s in all runs. Note, however, that the 
results are not sensitive to these values. 
The magnetorotation evolution 
of the neutron star depends on the field strength and its distribution 
in the crust. 
Our models at the beginning of the accretion phase are produced 
according to the framework discussed in \cite{urpin98mn}
for wind accretion rate $10^{-14}$ $M_{\odot}$/yr.   
The  magnetic field at the neutron star birth is assumed to 
be confined to the outer layers of the crust with densities 
$\rho \leq \rho_0~ 10^{13}$  g/cm$^3$ but this 
choice does not influence our main conclusions. Calculations 
are made for two values of the polar magnetic field at the 
neutron star birth, $B_p(0) = 3 \times 10^{13}$ and $6 \times 
10^{12}$G. This magnetic field processed in the course of phases 
I, II, and III then was used as the initial magnetic configuration 
for the phase IV. Our calculations of phase IV stop when the
magnetic field of a neutron star approaches the value $B_p = 
10^8$ G because, at the present, it is difficult to detect pulsars 
with $B_p < 10^8$ G.

  
In Fig.~1, we plot the evolutionary tracks of a neutron
star for three different accretion rates during the enhanced
accretion phase, $\dot{M} = 2.2 \cdot 10^{-9}$ (solid line), 
$4 \cdot 10^{-9}$ (dashed), and $5 \cdot 10^{-10}$ $M_{\odot}$/yr 
(dot-dashed). The critical temperature for the neutron $^3P_2$ 
gap is $10^9$ K and the magnetic field at the beginning of 
enhanced accretion is $8 \cdot 10^{11}$ G as it follows from
our choice of the initial condition for phase IV. After accretion
starts, the neutron star moves rapidly to the corresponding 
spin-up line and  after $9.8 \cdot 10^5$ yrs reaches it. In its 
further evolution, the star 
slides down the spin-up line until the balance is  maintained 
between the spin-up and rate of the field decay. However, 
this balance cannot be maintained for a long time because 
the magnetic field will eventually decay and  become too weak for that.
Due to the field decay, the Alfv\'en radius becomes small 
where the accretion flow interacts with a magnetosphere
and, therefore, the amount of angular momentum transferred to 
the neutron star is small as well. As a result, the neutron 
star begins to depart from the spin-up line. This behavior 
is typical for all considered models and  can be clarified
by Eq.~(7). The rate of angular momentum transfer is 
proportional to $B_p^{2/7}$ and, hence, it decreases with 
decreasing $B_p$. On the contrary, the rate of field decay 
does not depend on the magnetic field and, at a certain moment, 
the rate of angular momentum transfer should become smaller 
than the rate of field decay. Starting from this moment, the 
star cannot slide down the spin-up line anymore, Departures 
from the spin-up line become significant, however, at different 
$P$ for different accretion rates. A further evolution leads
to a formation of the pulsar with low magnetic field and
relatively long spin period. We define the minimum period, 
$P_m$, that the star can reach at a given accretion rate as 
the period that a neutron star has at $B_p = 10^8$ G, and we 
stop calculations at that time. For models considered in
Fig.~1, $P_m$ is equal to 
6.5 ms for the 
accretion rate of $2.2 \cdot 10^{-9}$ $M_{\odot}/$yr (solid line in Fig.~1).
The minimum period is determined by different mechanisms 
of field decay at different accretion rates. At a low accretion rate, 
$\dot{M} < 2.2 \times 10^{-9}$ $M_{\odot}/$yr, a decrease 
of the magnetic field is determined mainly by the ohmic 
dissipation (the second term on the r.h.s. of Eq,~(4)). In 
this case, the minimum period $P_m$ decreases if the accretion 
rate increases. At a high accretion rate, $\dot{M} > 2.2 
\times 10^{-9}$ $M_{\odot}/$yr, the advective term in 
Eq.~(4) (the first term on the r.h.s.) is greater than the 
dissipative one, and the accretion flow pushes the magnetic 
field into deep crustal layers. This submergence of the 
magnetic field leads to its rapid dissipation since we assume 
that the neutron star core is superconductive and the field 
cannot penetrate into the core. Therefore, the narrow transition 
layer is formed near the crust-core boundary and the field 
decays rapidly in this layer because of its relatively small 
thickness. The field decay is extremely fast in this case 
and its rate exceeds the rate of angular momentum transfer 
even if the magnetic field is relatively high. An increase 
of the accretion rate leads to more and more rapid decay of 
the field and, therefore, departures from the spin-up line 
can manifest itself even at higher magnetic fields. As a 
result, the minimum period increases with the increasing
accretion rate. Hence, the accretion rate exists at which the 
neutron star slides down the spin-up line during the longest 
time and, during this sliding down, the neutron star reaches 
the minimal value of minimum spin periods. If the accretion 
rate is higher or low than this value, the neutron star span 
up by accretion will rotate slower. For the model with the 
APR EOS, represented in Fig.~1, the minimal value of $P_m$ 
is $\approx 6.5$ ms and the corresponding accretion rate is 
$\approx 2 \times 10^{-9}$ $M_{\odot}/$yr. Note that the 
results are qualitatively the same for the neutron star 
models based on other EOS but the minimum periods have other 
values.

\begin{figure}
\includegraphics[width=9.0cm]{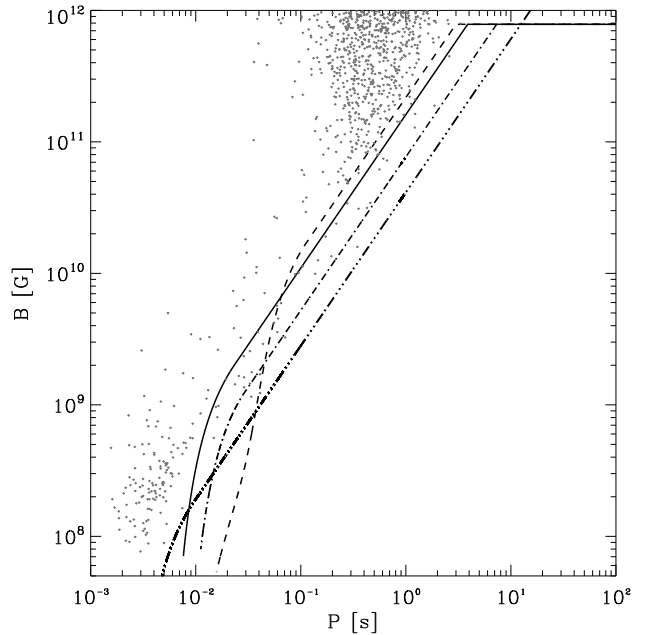}
\caption{Evolutionary tracks of pulsars in the $B$-$P$
plane for few accretion rates, $\dot{M} = 2.2 \times 10^{-9}$ 
(solid), $4 \times 10^{-9}$ (dashed), and $5 \cdot 10^{-10}$   
(dash-and-dotted) $M_{\odot}/$yr. The magnetic field at the
beginning of enhanced accretion is $8 \cdot 10^{11}$ G.
The long-dashed represent the same model of the solid line
but for a stiffer EOS (mean-field) for 1.4 $M_{\odot}$.}
\end{figure}

In Fig.~2, we plot the minimum period defined above, $P_m$, 
as a function of the accretion rate for the neutron star models 
based on the APR EOS and mean-field EOS.  In particular the solid curve represents 
a neutron star with
a  critical temperature for the $^3P_2$ gap of
$10^9$K,  (model ``a" in  \cite{minipage} see also \cite{elga96}) , 
and the magnetic field of $8 \cdot 10^{11}$ G when accretion starts. 
The  dot-dashed line represents the same
model of the solid line but with $T_c=10^{10}$ K, (gap model ``c"). 
The dashed line instead correspond 
to the same model of the dot-dashed line, but 
with a smaller magnetic field at the beginning of the phase
IV, $B_p(0) =1.6 \times 10^{11}$.
On  the other hand, the long-dashed and triple-dot-dashed line represent 
the same model of the solid-line but for 
a stiffer EOS (mean-field) for 1.4 $M_\odot$ and 1.6 $M_\odot$ solar masses, 
respectively.
In this latter case the minimum period is as low as 4 ms.

By comparing the dot-dashed line and the dashed line, we 
notice that the final spin period  clearly depends on the 
initial field strength being higher for a smaller magnetic
field, as it is expected. 
On the other hand by comparing the solid-line and the dot-dashed line
we observe that the spin period does not depend on the superconducting 
gap at the beginning of the accretion phase if the accretion rate 
is not very high, $\dot{M} \leq  2.2 \times 10^{-9}$ $M_{\odot}/$yr.
However, this is not the case if accretion is heavy and
$\dot{M} \geq  2.2 \times 10^{-9}$ $M_{\odot}/$yr 
the resulting minimal period is longer for higher $T_c$.

We can conclude that the minimal spin period is essentially 
determined by the physical property of the neutron star matter.
In particular we argue that $P_m$ depend on the EOS and can be 
significantly shorter than $\approx 7$ ms if an EOS stiffer than 
the APR one is used \citep{urpin98aa,urpin98mn}. {For example,
the minimal spin period is $\sim 4$ ms for the EOS by \citep{gl}.
Therefore, measurements of the minimal spin period of neutron 
stars processed in low-mass binaries can provide information 
regarding the EOS. Our study shows that stiff EOSs are in a better
agreement with observational data.
}

\begin{figure}
\includegraphics[width=9.0cm]{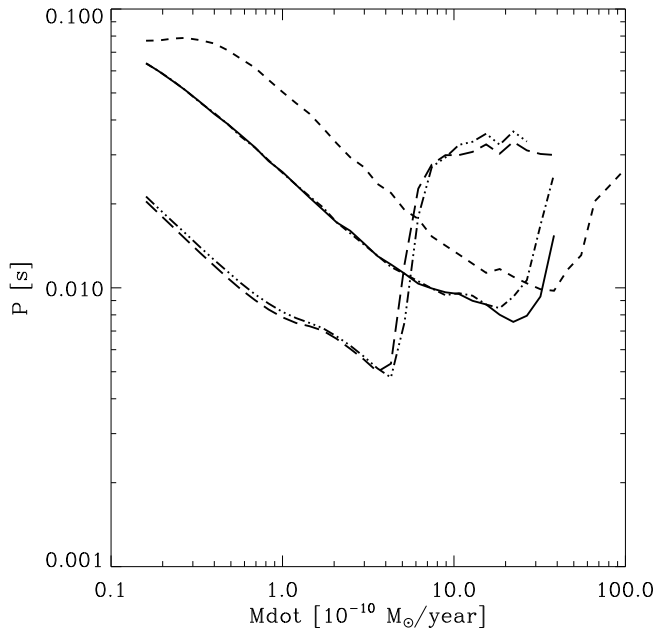}
\caption{The minimum period, $P_m$ that the neutron star
can reach by accretion as a function of the accretion rate.
The solid line corresponds to a neutron star with APS EOS and the 
magnetic field at the beginning of
enhanced accretion $8 \times 10^{11}$ G and a neutron $^3P_2$ gap 
with $T_c=10^9$K. The dot-dashed line is the same model of the solid
line but with $T_c = 10^{10}$ K. The dashed line 
represents the same model of the dot-dashed 
but for the initial field of  $1.6 \times 10^{11}$ G.
The long-dashed and triple-dot-dashed line represent the same model 
of the solid-line but for 
a stiffer EOS (mean-field) for 1.4 $M_\odot$ and 1.6 $M_\odot$ solar 
masses, respectively.
}

\end{figure}

\section{Discussion}

We have considered a combined magnetic, thermal, and rotational 
evolution of the accreting neutron star based on different EOSs. 
It turns out that the accreting neutron star cannot spin up to 
a very short period even if the accretion rate is high and 
accretion lasts sufficiently long time. The reason for this 
is a strong coupling between the rotational, magnetic, and 
thermal evolution during the accretion phase. At a given 
accretion rate, there exists the minimum period, $P_m$, that 
the neutron star cannot overcome. This limit is caused by the 
fact that accretion-driven decay of the magnetic field leads 
to a rapid decrease of the angular momentum transfer. At some 
stage, the rate of the angular momentum transfer becomes so 
low that the neutron star cannot maintain a balance with the 
rate of field decay in order to slide down the spin-up line. 
Since the field decay is faster, the star leaves the spin-up 
line and evolves into the region of low-magnetic and relatively 
long-periodic pulsars in the $B$-$P$ plane. 

{The point where a neutron star begins to depart from the 
spin-up line can be qualitatively estimated by comparing the
rate of angular momentum transfer with the rate of field decay.
The inverse timescale of spin-up is given by Eq.~(7), $\dot{P}/P 
\sim - Pj/2 \pi I$. Since $j \sim \dot{M} \Omega_K(R_A) R_A^2 
\sim \dot{M} \sqrt{GM} R_A^{1/2}$ during the enhanced accretion 
phase, we obtain
\begin{equation}
\frac{\dot{P}}{P} \sim - \frac{\dot{M} P}{2 \pi I} \sqrt{GM} R_A^{1/2}
\sim - \frac{\Omega_K(R_A)}{\Omega} \cdot \frac{\dot{M} R_A^2}{I}.
\end{equation}
On the other hand, the inverse timescale of the magnetic field decay
can be estimated as $\dot{B}/B \sim - 1/ \Delta t$ where $\Delta
t$ is the time from the beginning of the enhanced accretion (see, 
e.g., Urpin \& Geppert 1995). The field decay is faster than 
spin-up if $\dot{B}/B > \dot{P}/P$, or
\begin{equation}
\Delta M R_A^2 > I (\Omega / \Omega_K(R_A)),
\end{equation}
where $\Delta M = \dot{M} \Delta t$ is the accreted mass. 
Generally, departures from the spin-up line depend on the magnetic 
field and the mass and radius of a neutron star and, hence, on the
equation of state of nuclear matter. Neutron stars with a soft EOS 
(with smaller $R$ at given $M$) need to accrete a smaller mass to 
leave the spin-up line. On the contrary, stars with a stiff EOS 
(greater $R$ at given $M$) leave this line only if $\Delta M$ is 
sufficiently large.
}

{In this paper, we define $P_m$ as the period reached by 
a neutron star when its magnetic field becomes equal to $10^8$ 
G. This definition is rather arbitrary and we introduce it mainly
in order to stop calculations when the rotational evolution 
becomes very slow. In fact, the spin period of a neutron star 
continues to decrease slowly after it reaches the value $P_m$. 
However, this decrease turns out to be very slow and it becomes 
slower during the further evolution. For example, 
$P_m$  is equal $\sim 6$ ms if we stop calculations when the 
magnetic field reaches the value $10^7$ G in comparison with 
$P_m \approx 7$ ms obtained for $B=10^8$ G in Fig.2. Besides, 
detection of the neutron stars with $B < 10^8$ G is a very 
complicated problem for the present telescopes. Therefore, we 
stop calculations when $B < 10^8$ G  and restrict our discussion 
by the properties of neutron stars that can be observed at the 
present. 
}

{The main reason for this relatively high minimal period of 
accreting pulsars is the rapid decrease of the magnetic field which 
leads to a low value of $R_A$. Transfer of the angular momentum 
becomes too slow at small $R_A$. If the accretion flow forms the 
Keplerian disc around a neutron star, the rate of angular momentum 
transfer can be estimated as $\dot{M} \Omega_K(R_A) R_A^2$. Note 
that this estimate is valid only if $R_A \gg R$. For the considered 
accretion rate, this condition is satisfied if the magnetic field 
of a neutron star is greater than $\sim 10^7$ G. If the magnetic 
field is very weak ($B_p < 10^7$ G) then $R_A \leq R$ and the rate 
of angular momentum transfer is of the order of $\dot{M} \Omega_K(R) 
R^2$. Since $\Omega_K(R) \propto R^{-3/2}$, the angular momentum 
transport becomes very slow in this case. Despite the inner disc 
radius cannot be smaller than the radius of the inner marginally 
stable orbit, the rate of angular momentum transport turns out to 
be low even if the accretion disc is extended to this orbit.  
Therefore, the further spin evolution of a neutron star is very
slow and, in order to reach short periods, enhanced accretion 
should last extremely long time. 
}

As it was noted, the minimum period, $P_m$, is determined by the 
accretion rate and EOS of nuclear matter. However, this minimum 
period cannot be very short even if the accretion rate is high. 
A high accretion rate leads to a very rapid decrease of the
magnetic field. In this case, the magnetic evolution in 
the crust is dominated by advection: the accretion flow 
pushes the magnetic field into the deep layers where the field 
dissipates rapidly in the narrow transition region near the 
superconductive core. A fast submergence of the magnetic field 
leads to a rapid decrease of the rate of the angular momentum 
transfer and substantial departures from the spin-up line.
Due to this, the neutron star does not spend much time, sliding 
down the spin-up line, and can spin up only to relatively long 
periods. As a result, there exists the accretion rate at which 
$P_m$ reaches the minimal value for a given EOS. The minimal 
value of $P_m$ for the considered model with the APR EOS is 
$\approx 7$ ms and the neutron star reaches it if the accretion 
rate is $\approx  2 \times 10^{-9}$ $M_{\odot}/$yr. If the 
accretion rate is higher or lower than this value, the neutron 
star can spin up only to a longer period. This limit of the
spin period is determined by the EOS and can differ from the
value $\sim 7$ for other EOS. Note that the limit is smaller
for stiff EOSs and greater for soft EOSs. {For example,
the minimal period for a neutron star with the EOS by \citep{gl}
is $\sim 4$ ms and this period can be reached if the accretion
rate is $\sim 4 \times 10^{-10}$ $M_{\odot}/$yr. Note that, for
a stiffer EOS, the neutron star reaches the minimal period 
at lower accretion rate.   
}

The shortest spin period detected up to date for PSR J1748-244
is 1.4 ms \cite{hess06}. The minimal spin periods 
attainable for neutron stars and determined by balance of 
the gravity and centrifugal forces are well below this value 
for most of the proposed EOSs of a nuclear matter.
A discrepancy between the minimal possible spin
period of neutron stars and the shortest detected period can
be caused by the mechanism considered in this paper. 
{The calculated values of a minimal period, $\sim 7$ ms and 
$\sim 4$ ms, clearly point out that the EOS should be stiffer 
than that by \cite{apr} and, likely,  even stiffer than that 
by \citep{gl}}.

Our calculations show that formation of shortly periodic 
pulsars (like millisecond pulsars) is possible only if the 
accretion rate belongs to some not very wide range around
$\dot{M} \sim  2 \times 10^{-9}$ $M_{\odot}/$yr for APS EOS. If the 
accretion rate is lower than this value, a pulsar in the
binary system has no chance to become a millisecond one 
even if accretion lasts very long. Note that some
observational indications (for instance, a distribution 
of pulsars in the $B$-$P$ plane) also point out on the 
existence of the minimal accretion rate that can lead to 
a formation of millisecond pulsars \citep{pana}.

\vspace{0.5cm}
\noindent
{\it Acknowledgments.} We would like to thank Dany Page for kindly providing us with the mean-field EOS used in the calculation. 



\end{document}